\documentclass[12pt, a4paper]{article}
\pdfoutput=1

\usepackage{jcappub}
\usepackage{graphicx}
\usepackage{amsmath,amssymb}
\usepackage{bm}
\usepackage{color}
\usepackage{tensor}
\usepackage{hyperref}
\usepackage[all]{hypcap}
\usepackage{slashed}
\usepackage{multirow}
\usepackage{empheq}
\usepackage{ulem}
\usepackage{comment}

\usepackage{myterms}

\newcommand{\eff}{{\rm eff}}
\newcommand{\vis}{{\rm vis}}
\newcommand{\cadi}{c_{\rm adi}}
\newcommand{\ceff}{c_{\rm eff}}
\newcommand{\cvis}{c_{\rm vis}}

\newcommand{\xvec}{{\bm x}}

\newcommand{\rvec}{{\bm r}}

\newcommand{\kvec}{{\bm k}}

\newcommand{\khat}{\hat{\kvec}}
\newcommand{\pvec}{{\bm p}}

\newcommand{\qvec}{{\bm q}}

\newcommand{\qhat}{\hat{\qvec}}

\newcommand{\ebar}{\epsilon}

\newcommand{\drho}{\delta \mkern-1.5mu \rho}

\newcommand{\dN}{\delta \mkern-1.5mu N}
\newcommand{\dP}{\delta \mkern-1.5mu P}
\newcommand{\dQ}{\delta \mkern-0.5mu Q}
\newcommand{\dS}{\delta \mkern-1.5mu S}
\newcommand{\dC}{\delta \mkern-0.5mu C}
\newcommand{\dPtilde}{\delta \mkern-1.5mu \tilde{P}}
\newcommand{\dQtilde}{\delta \mkern-0.5mu \tilde{Q}}
\newcommand{\dStilde}{\delta \mkern-1.5mu \tilde{S}}

\title{Sound Speed and Viscosity of Semi-Relativistic Relic Neutrinos}

\author[a,b]{Lawrence Krauss}
\author[c]{and Andrew J. Long}

\affiliation[a]{Physics Department and School of Earth and Space Exploration, Arizona State University, Tempe, Arizona 85287, USA.}
\affiliation[b]{Research School of Astronomy and Astrophysics, Mt. Stromlo Observatory, Australia National University, Canberra, Australia 2611.}
\affiliation[c]{Kavli Institute for Cosmological Physics, The University of Chicago, Chicago, Illinois 60637, USA.}

\emailAdd{krauss@asu.edu}
\emailAdd{andrewjlong@kicp.uchicago.edu}

\date{\today}

\abstract{
Generalized fluid equations, using sound speed $c_{\rm eff}^2$ and viscosity $c_{\rm vis}^2$ as effective parameters, provide a convenient phenomenological formalism for testing the relic neutrino ``null hypothesis,'' {\it i.e.} that that neutrinos are relativistic and free-streaming prior to recombination.  
In this work, we relax the relativistic assumption and ask ``to what extent can the generalized fluid equations accommodate finite neutrino mass?''  
We consider both the mass of active neutrinos, which are largely still relativistic at recombination $m^2 / T^2 \sim 0.2$, and the effect of a semi-relativistic sterile component.  While there is no one-to-one mapping between mass/mixing parameters and $c_{\rm eff}^2$ and $c_{\rm vis}^2$, we demonstrate that the existence of a neutrino mass could induce a bias to measurements of $c_{\rm eff}^2$ and $c_{\rm vis}^2$ at the level of $0.01 m^2 / T^2 \sim 10^{-3}$.  
}


\begin{document}
\maketitle

\setlength{\parindent}{20pt}
\setlength{\parskip}{2.5ex}

\section{Introduction}\label{sec:Introduction}

Precision measurements of the cosmic microwave background (CMB) and large scale structure (LSS) are providing a wealth of information about the early universe and its constituents.  
This information is particularly valuable in the neutrino sector where a number of fundamental questions have yet to be answered:  What is the absolute neutrino mass scale?  Are some neutrinos sterile?  Do neutrinos self-interact through a long range force?  
The next-generation of CMB and LSS experiments will bring dramatic improvements in sensitivity and the promise of new insight into the physics of neutrinos \cite{Abazajian:2013oma}.  

To address the questions listed above in a model-independent way, it is customary to use phenomenological parameters.  
These parameters are introduced ``by hand'' into the equations of motion (Einstein or Boltzmann equations).  
They are not defined by any underlying fundamental parameters, such as Lagrangian couplings or masses.  

The most familiar phenomenological parameters are the effective number of neutrino species $N_{\rm eff}$ and the total neutrino mass $\sum m_{\nu}$.  
Since the relic neutrinos are decoupled at the time of recombination and structure formation, their effect on the CMB and LSS are only gravitational.  
Thus, the phenomenological parameters encode how much the neutrinos contribute to the energy densities (see \cite{Abazajian:2013oma,Mangano:2005cc} for notation)
\begin{align}\label{eq:Neff_and_summnu}
	\rho_{\rm rad} = \rho_{\gamma} + N_{\eff} \frac{7}{8} \left( \frac{4}{11} \right)^{4/3} \rho_{\gamma}
	\qquad \text{and} \qquad
	\Omega_{\nu} h^2 = \frac{\sum m_{\nu}}{93.1 \eV} 
	\per
\end{align}
Since $N_{\rm eff}$ and $\sum m_{\nu}$ are not defined from fundamental parameters, there does not necessarily exist a one-to-one mapping from any specific microphysical model onto the parameters $(N_{\rm eff}, \sum m_{\nu})$.  
Rather, the phenomenological parameters are most useful as a test of the ``null hypothesis.''  
A combination of the concordance cosmology and Standard Model of particle physics predicts $N_{\rm eff} = 3.046$ and $\sum m_{\nu} = m_1 + m_2 + m_3 > 0.05 \eV$ where $m_i$ are the three neutrino mass eigenvalues.  
Measurements compiled by the Planck collaboration \cite{Ade:2015xua} ({\it Planck}, TT + lensing + ext), 
\begin{align}\label{eq:measure_Neff}
	N_{\rm eff} \simeq 3.15 \pm 0.40
	\qquad \text{and} \qquad
	\sum m_{\nu} < 0.234 \eV \quad \text{at 95\% CL}
	\com
\end{align}
are consistent with the null hypothesis.  

Two additional phenomenological parameters affect the evolution of neutrino density inhomogeneities.  
These are the effective sound speed $\ceff^2$ and viscosity $\cvis^2$ \cite{Hu:1998kj,Hu:1998tk}.  
The effective sound speed sets the sound horizon, which in turn controls the growth of neutrino density perturbations, and the viscosity parameter leads to an anisotropic stress and the damping of neutrino density perturbations.  
(See Refs.~\cite{Audren:2014lsa} for a discussion of these effects on the CMB.)  
Once again, the phenomenological parameters provide a model-independent formalism to test the null hypothesis:  if the relic neutrinos are relativistic and free-streaming then one expects $\ceff^2 = 1/3$ and $\cvis^2 = 1/3$.  
The Planck collaboration furnishes the measurements \cite{Ade:2015xua} ({\it Planck}, TT, TE, EE + lowP + BAO)
\begin{align}\label{eq:measure_csq}
	\ceff^2 \simeq 0.3242 \pm 0.0059
	\qquad \text{and} \qquad
	\cvis^2 \simeq 0.331 \pm 0.037
	\com
\end{align}
which are consistent with the null hypothesis.  

As measurements of the four phenomenological parameters improve with the next generation of CMB and LSS experiments, we must be mindful of any deviation from the null hypothesis, as this would indicate the presence of new physics.  
In order to probe the nature of the new physics, we must understand how a specific microphysical model maps onto the phenomenological parameters.  
For instance, many studies have investigated how {\rm eV}-scale sterile neutrinos (motivated in part by the short baseline and reactor anomalies \cite{Abazajian:2012ys, Giunti:2015wnd}) manifest themselves in the CMB and LSS (for one such recent paper see \rref{Costanzi:2014tna}).  
This provides a mapping from the sterile mass and abundance to $N_{\rm eff}$ and $\sum m_{\nu}$.  
We seek to extend that correspondence to the perturbation parameters $c_{\eff}^2$ and $c_{\vis}^2$.  

In \sref{sec:effective} we study the formalism (generalized fluid equations) in which the phenomenological parameters $\ceff^2$ and $\cvis^2$ arise.  
While this formalism is convenient for testing the null hypothesis, we will see that it cannot generally accommodate realistic deviations from the null hypothesis.  
Specifically, if the neutrinos are assumed to be free-streaming but allowed to be semi-relativistic (such is the case for sterile neutrinos) then the fluid equations describing their evolution cannot be mapped onto the generalized fluid equations.  
In \sref{sec:Deviations} we estimate the dependence of $\ceff^2$ and $\cvis^2$ on neutrino mass, and we calculate the predicted deviations from the null hypothesis, $1/3 - \ceff^2$ and $1/3 - \cvis^2$, for a model of sterile neutrinos that saturates the Planck limits in \eref{eq:measure_Neff}.  
We summarize our results and discuss directions for future work in \sref{sec:Discussion}.  
The main paper is accompanied by \aref{app:FluidEqns}, where we derive the fluid equations for a free-streaming species from the Boltzmann hierarchy.  
\aref{app:FD_dist} contains formulas relevant to a semi-relativistic Fermi-Dirac phase space distribution.  

\section{Fluid Equations for Semi-Relativistic, Free-Streaming Neutrinos}\label{sec:effective}

We are interested in the background of relic neutrinos at temperatures $T \lesssim 3 \MeV$ when the weak interactions have frozen out and the neutrinos are decoupled from the plasma.  
Standard Model neutrinos experience no additional interactions, and they are free-streaming.  
Later we will extend the model to include {\rm eV}-scale sterile neutrinos, which are also assumed to be free-streaming.  

\begin{figure}[t]
\hspace{0pt}
\vspace{-0.0cm}
\begin{center}
\includegraphics[width=0.48\textwidth]{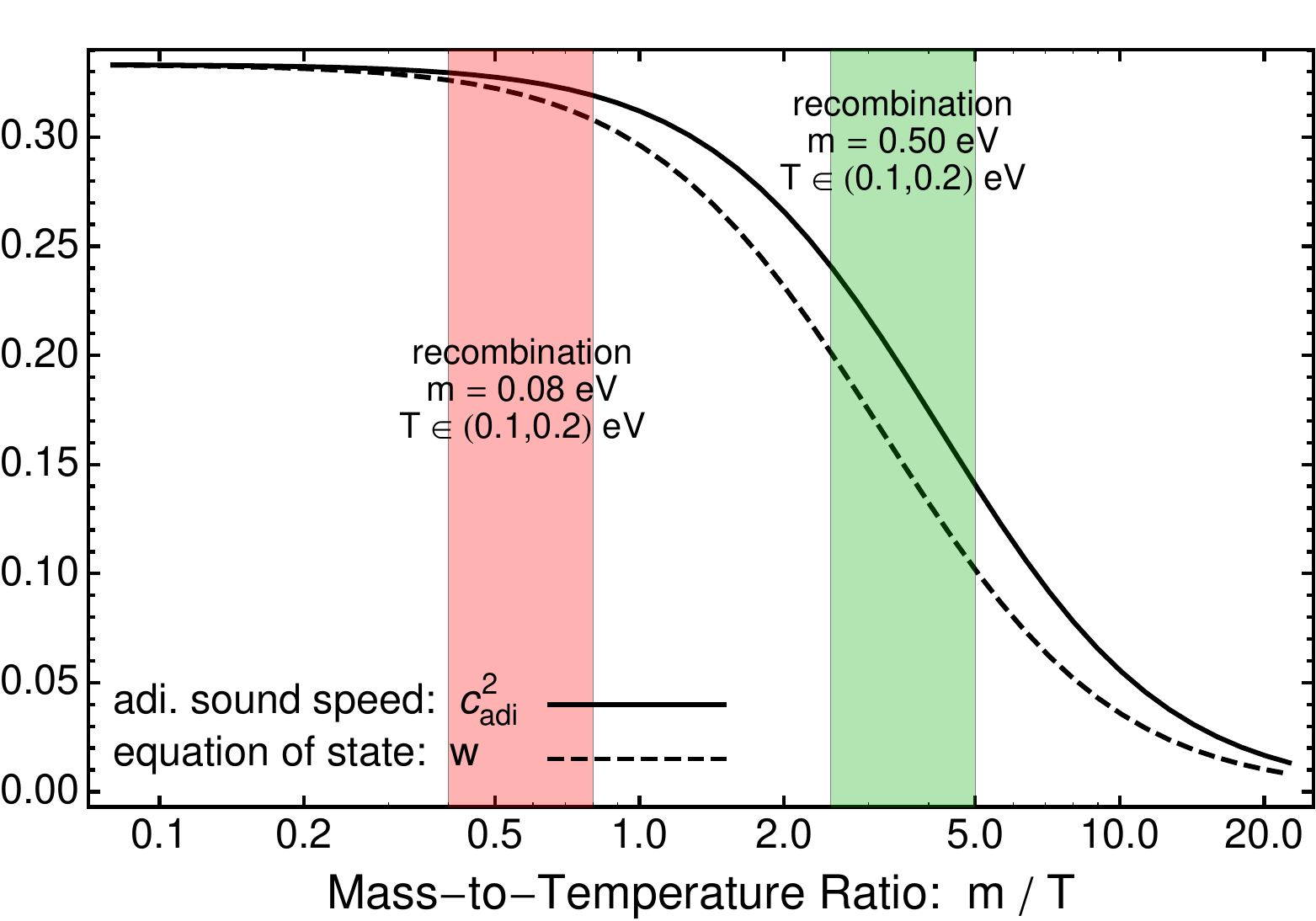} \hfill
\includegraphics[width=0.48\textwidth]{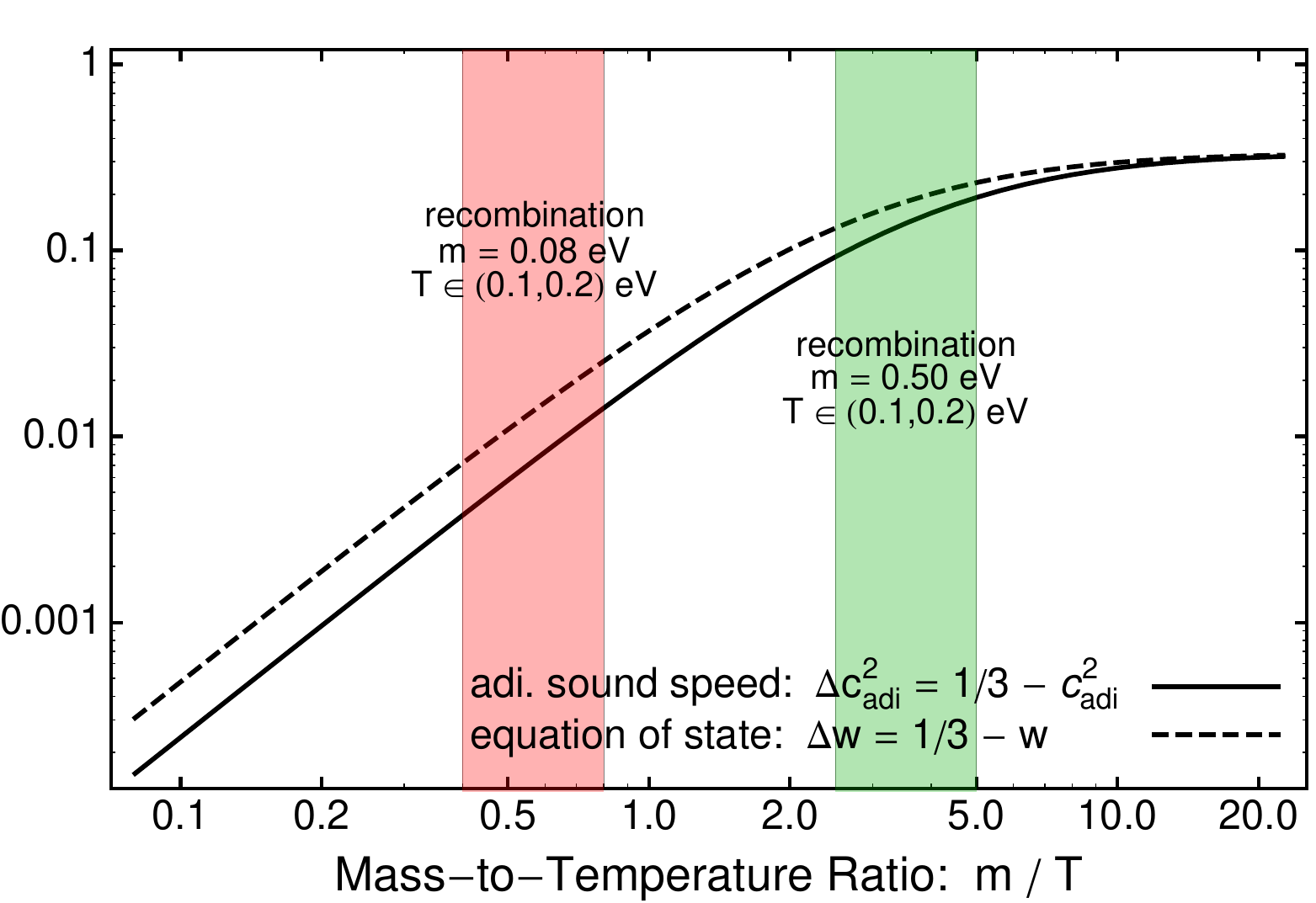} 
\end{center}
\caption{\label{fig:w_cadi2}  
The equation of state $w$ (solid) and squared adiabatic sound speed $c_{\rm adi}^2$ (dashed) for a Fermi-Dirac distribution of relic neutrinos with mass $m$ and temperature $T$.  The shaded regions indicates the epoch of recombination, $T_{\rm rec} \approx 0.1-0.2 \eV$, for $m = 0.08 \eV$ and $0.5 \eV$.  
}
\end{figure}

To leading order, the medium is homogeneous with energy density $\bar{\rho}(\tau)$ and pressure $\bar{P}(\tau)$.  
The corresponding equation of state and adiabatic sound speed are $w = \bar{P} / \bar{\rho}$ and $c_{\rm adi}^2 = \dot{\bar{P}} / \dot{\bar{\rho}}$, where the dot indicates differentiation with respect to conformal time $\tau$.  
After decoupling the neutrino background maintains its Fermi-Dirac distribution with temperature $T$.  
Using the notation established in \aref{app:FluidEqns} we calculate $w$ and $c_{\rm adi}^2$ in \aref{app:FD_dist} and present the result in \fref{fig:w_cadi2}.  
For simplicity we assume that the neutrino spectrum is in the degenerate regime, and the common neutrino mass is $m \approx (1/3) \sum m_{\nu}$.  
Initially the neutrino temperature is high, $m \ll T$, and $w, c_{\rm adi}^2 \approx 1/3$ for the relativistic neutrinos.  
As the temperature is lowered, the deviations $\Delta w = 1/3 - w$ and $\Delta c_{\rm adi}^2 = 1/3 - c_{\rm adi}^2$ start to grow as the neutrinos become semi-relativistic.  
For a Fermi-Dirac distribution we find
\begin{align}\label{eq:Dw_Dcadi2}
	\Delta \cadi^2 \approx \frac{\Delta w}{2} \approx \frac{5}{21 \pi^2} \frac{m^2}{T^2} 
\end{align}
for small $m / T$.  
The anomalously small prefactor, $5/21\pi^2 \simeq 0.02$, invalidates the naive dimensional analysis prediction $\Delta \cadi^2 \sim m^2 / T^2$.  

During recombination, the photon temperature is $T_{\gamma} \approx 0.2 - 0.3 \eV$ and the neutrino temperature is smaller by a factor of $(4/11)^{1/3}$, which corresponds to $T_{\rm rec} \approx 0.1 - 0.2 \eV$.  
Taking a fiducial neutrino mass of $m = 0.08 \eV$, which saturates the Planck bound in  \eref{eq:measure_Neff}, the deviations fall into the range $0.4 \% \lesssim \Delta c_{\rm adi}^2 \lesssim 2 \%$ at the time of recombination.  
For a heavier {\rm eV}-scale sterile neutrino, the deviation is $(10-20)\%$ (assuming the phase space distribution function is also Fermi-Dirac).  
The observation that $\Delta w / w \ll 1$ and $\Delta \cadi^2 / \cadi^2 \ll 1$ has two implications for our analysis.  
It indicates that we can study deviations from the relativistic relic neutrino background by perturbing in the small quantities $\Delta w / w$ and $\Delta \cadi^2 / \cadi^2$.  
Additionally it suggests that the effects of finite neutrino mass will be at most $\sim 0.02 m^2 / T^2$ in magnitude.  

Let us now consider perturbations to the homogenous Fermi-Dirac distribution.  
The details of this calculation appear in \aref{app:FluidEqns}.  
Since the inhomogenous phase space distribution function depends on both momentum $\qvec$ and position (or wavevector $\kvec$ in Fourier space), it is convenient to organize the perturbations into multipole moments with index $\ell$.  
Each moment of the phase space distribution function can be integrated over momentum $q = \abs{\qvec}$.  
It is possible to include additional factors of the momentum-to-energy ratio $q/\epsilon$ in the integrand.  
For the lowest order multipole moments ($\ell = 0,1,2$) one obtains the energy density contrast $\delta(\kvec,\tau)$, energy flux $\theta(\kvec,\tau)$, and anisotropic stress $\sigma(\kvec,\tau)$.  
Other combinations of $\ell$ and $q/\epsilon$ lead to a doubly-infinite tower of perturbation variables, shown in \tref{tab:perturbations}; see \erefs{app:moments_1}{app:moments_2} for detailed expressions.  
Specifically, $\Pi(\kvec,\tau)$ is the pressure perturbation.  
In the non-relativistic limit, $T/m \ll 1$, the perturbation variables constructed from additional factors of $q/\epsilon$ are suppressed by powers of $T/m$.  
In the relativistic limit, $T/m \gg 1$, additional factors of the momentum-to-energy ratio simplify $q/\epsilon \approx 1$ leading to 
\begin{align}\label{eq:pert_rel_limit}
	\tilde{\Pi} \approx \Pi \approx \delta / 3
	\ , \quad 
	\tilde{\theta} \approx \theta
	\ , \quad \text{and} \qquad
	\tilde{\sigma} \approx \sigma 
	\per
\end{align}

\begin{table}[t]
\begin{center}
\begin{tabular}{c||c|c|c|c|c|c}
$\ell$ & $\epsilon$ & $q$ & $q^2 / \epsilon$ & $q^3 / \epsilon^2$ & $q^4 / \epsilon^3$ & $\cdots$ \\
\hline
0 & $\delta$ & {} & $\Pi$ & {} & $\tilde{\Pi}$ & $\cdots$ \\
1 & {} & $\theta$ & {} & $\tilde{\theta}$ & {} & $\cdots$ \\
2 & {} & {} & $\sigma$ & {} & $\tilde{\sigma}$ & $\cdots$ \\
3 & {} & {} & {} & $\chi$ & {} & $\cdots$ \\
$\vdots$ & $\vdots$ & $\vdots$ & $\vdots$ & $\vdots$ & $\vdots$ & $\ddots$
\end{tabular}
\end{center}
\caption{\label{tab:perturbations}The perturbation variables discussed in the text can be organized according the multipole moment $\ell$ of the phase space distribution function from which they were calculated, and the factors of energy $\epsilon$ and momentum $q$ that were included in the momentum integral.  Numerical factors of $1/3$, etc., are not shown; see \erefs{app:moments_1}{app:moments_2} for detailed expressions.  }  
\end{table}%

The evolution of perturbations in a system of freely streaming particles is described by the collisionless Boltzmann equation \cite{Ma:1995ey}.  
Upon performing the multipole expansion described above, the Boltzmann equation yields a hierarchy of coupled first order differential equations describing the evolution of each moment.  
See \aref{app:FluidEqns} for details of this calculation.  
Focusing on the first few multipole moments, we perform the momentum integrals to obtain
\begin{subequations}\label{eq:Fluid_Eqns_massive}
\begin{align}
	\dot{\delta} & 
	= - (1+w) \Bigl( \theta + \frac{1}{2} \dot{h} \Bigr) + 3 \frac{\dot{a}}{a} \Bigl( w \delta - \Pi \Bigr) \\
	\dot{\theta} & 
	= - 3 \frac{\dot{a}}{a} \Bigl( \frac{1}{3}- c_{\rm adi}^2 \Bigr) \theta + k^2 \frac{\Pi}{1 + w} - k^2 \sigma \\
	\dot{\sigma} & 
	= \frac{4}{15} \theta + (3 c_{\rm adi}^2) \frac{2}{15} \bigl( \dot{h} + 6 \dot{\eta} \bigr) - \frac{3}{5} k \chi \nn
	& \quad - 3 \frac{\dot{a}}{a} \Bigl( \frac{1}{3} - c_{\rm adi}^2 \Bigr) \sigma  + \frac{\dot{a}}{a} \bigl( \tilde{\sigma} - \sigma \bigr) + \frac{4}{15} \bigl( \tilde{\theta} - \theta \bigr) \\
	\dot{\Pi} & 
	= - 3 \frac{\dot{a}}{a} \Bigl( \frac{1}{3} - w \Bigr) \Pi + \frac{\dot{a}}{a} \bigl( \tilde{\Pi} - \Pi \bigr) - \frac{1}{3} (1+w) \tilde{\theta} - \frac{1}{6} \dot{h} \bigl( 5 w - \tilde{w} \bigr) 
	\com
\end{align}
\end{subequations}
which we call the collisionless fluid equations\footnote{This name is something of an oxymoron.  In a perfect fluid, collisions occur frequently and tend to isotropize the perturbations.  This enforces a vanishing of the anisotropic stress $\sigma$ and higher multipole moments.  One should view \eref{eq:Fluid_Eqns_massive} as the analog of the fluid equations for a free-streaming species.  }.  
We are working in the synchronous gauge where the metric perturbations are denoted as $h(\kvec,\tau)$ and $\eta(\kvec,\tau)$, and their evolution is given by Einstein's equations.  
The equations for $\dot{\delta}$ and $\dot{\theta}$ are the familiar continuity and Euler equations\footnote{These can also be derived from the conservation of stress-energy \cite{Ma:1995ey}.  }.  
Note that $k = \abs{\kvec}$ and $a(\tau)$ is the FRW scale factor.  
The parameter $\tilde{w}$ is the pseudo-equation of state, defined in the appendix.  
The equation for $\dot{\sigma}$ depends on the next moment ($\ell=3$) in the multipole expansion $\chi(\kvec,\tau)$.  
This is the familiar result for the Boltzmann hierarchy:  the evolution of lower-order multipole moments depends on the higher-order moments.  
In addition, the equations for $\dot{\sigma}$ and $\dot{\Pi}$ also depend on the tilde'd variables $\tilde{\theta}$, $\tilde{\sigma}$, and $\tilde{\Pi}$.  
Consequently the equations shown in \eref{eq:Fluid_Eqns_massive} do not form a closed system.  
However, we are only interested in comparing the form of these equations with the generalized fluid equations below, and for that purpose we do not require the rest of the hierarchy.  

In the ultra-relativistic regime, $m \ll T$, we can approximate $w \approx c_{\rm adi}^2 \approx 1/3$.  
Additionally, the perturbation variables reduce as in \eref{eq:pert_rel_limit}.  
Then \eref{eq:Fluid_Eqns_massive} becomes
\begin{subequations}\label{eq:Fluid_Eqns_massless}
\begin{align}
	\dot{\delta} & 
	= - \frac{4}{3} \Bigl( \theta + \frac{1}{2} \dot{h} \Bigr) \\
	\dot{\theta} & 
	= \frac{1}{4} k^2 \delta - k^2 \sigma \\
	\dot{\sigma} & 
	= \frac{4}{15} \theta + \frac{2}{15} \bigl( \dot{h} + 6 \dot{\eta} \bigr) - \frac{3}{5} k \chi 
	\com
\end{align}
\end{subequations}
and $\Pi = \delta / 3$, which are the fluid equations for free-streaming, relativistic particles.  

A phenomenological generalization of the fluid equations was proposed in \rref{Hu:1998kj,Hu:1998tk}.  
By introducing the sound speed and the viscosity parameters, $c_{\eff}^2$ and $c_{\vis}^2$, one can write \cite{Archidiacono:2011gq}\footnote{In comparing with Eqs.~(2--4) of \rref{Archidiacono:2011gq}, note that $q_{\nu}(\kvec,\tau) = 4 \theta(\kvec,\tau) / (3k)$ and $\pi_{\nu}(\kvec,\tau) = 2 \sigma(\kvec,\tau)$ and $F_{\nu,3}(\kvec,\tau) = 2 \chi(\kvec,\tau)$ in the massless limit.  }
\begin{subequations}\label{eq:Fluid_Eqns_generalized}
\begin{align}
	\dot{\delta} & = - \frac{4}{3} \Bigl( \theta + \frac{1}{2} \dot{h} \Bigr) + 3 \frac{\dot{a}}{a} \Bigl( \frac{1}{3} - c_{\rm eff}^2 \Bigr) \delta + 12 \Bigl( \frac{\dot{a}}{a} \Bigr)^2 \Bigl( \frac{1}{3} - c_{\rm eff}^2 \Bigr) \frac{\theta}{k^2} \\
	\dot{\theta} & = -3 \frac{\dot{a}}{a} \Bigl( \frac{1}{3} - c_{\eff}^2 \Bigr) \theta + \frac{1}{4} (3c_{\eff}^2) k^2 \delta - k^2 \sigma \\
	\dot{\sigma} & = (3 c_{\vis}^2) \frac{4}{15} \theta + (3 c_{\vis}^2) \frac{2}{15} \bigl( \dot{h} + 6 \dot{\eta} \bigr) - \frac{3}{5} k \chi
	\com
\end{align}
\end{subequations}
which we call the generalized fluid equations (GFE).  
The rest of the Boltzmann hierarchy, {\it e.g.} the equation for $\dot{\chi}$, is unmodified.  
Relativistic and free-streaming neutrinos obey the fluid equations in \eref{eq:Fluid_Eqns_massless}, which corresponds to the limit $c_{\eff}^2 = c_{\vis}^2 = 1/3$ in the GFE.  
Therefore, measuring a deviation from $1/3$ would refute the ``null hypothesis,'' {\it i.e.} that the relic neutrinos are relativistic and free-streaming.  
A number of studies have investigated the effects of $\ceff^2$ and $\cvis^2$ on the cosmic microwave background \cite{Trotta:2004ty, DeBernardis:2008ys, Lesgourgues:2011rh, Smith:2011es, Diamanti:2012tg, Archidiacono:2012gv, Gerbino:2013ova, Archidiacono:2013fha, Archidiacono:2013lva, Sellentin:2014gaa, Audren:2014lsa, DiValentino:2016ikp}, and recently the Planck collaboration reported the measurements in \eref{eq:measure_csq} using a combination of CMB and BAO data.  
These measurements illustrate the utility of the generalized fluid equations for testing -- and thus far confirming -- the null hypothesis of relativistic and free-streaming neutrinos.  

However, it is not clear the extent to which the GFE is able to capture specific models when we relax the assumptions of relativistic free-streaming particles.  
For instance, it is often said that $(c_{\eff}^2 \, , \, c_{\vis}^2)=(1/3,0)$ corresponds to a relativistic perfect fluid, and therefore this limit has been used to model the effect of neutrino self-interactions \cite{Beacom:2004yd, Hannestad:2004qu, Bell:2005dr, Sawyer:2006ju, Basboll:2008fx, Oldengott:2014qra} (see also \cite{Archidiacono:2013dua}).  
However, while $c_{\vis}^2=0$ allows for solutions in which the anisotropic stress and higher moments vanish as in a perfect fluid, it also allows for solutions where they are nonzero and static, which is not the case for a perfect fluid.  
These criticisms were recently raised by Refs.~\cite{Shoji:2010hm,Sellentin:2014gaa,Oldengott:2014qra}.  

In this work, we consider the effect of finite neutrino mass either arising from the active neutrinos themselves or a heavier sterile neutrino component.  
This problem has been investigated recently in \rref{Audren:2014lsa} by numerically solving the Boltzmann hierarchy, and it was found that there is no clear degeneracy between neutrino mass and the sound speed parameters.  
Our goal is to develop an analytic understanding of this result while also deriving a parametric relationship between the parameters of the GFE and the neutrino mass.  

If we relax the assumption of relativistic neutrinos but maintain the assumption of free-streaming neutrinos, then the density perturbations satisfy the collisionless fluid equations of \eref{eq:Fluid_Eqns_massive}.  
Clearly it is not possible to put the GFE of \eref{eq:Fluid_Eqns_generalized} into the form of \eref{eq:Fluid_Eqns_massive} even with a judicious choice of the parameters $(\ceff^2 \, , \, \cvis^2)$; the equations have different structures.  
However, the neutrinos are still semi-relativistic at the time of recombination, see \fref{fig:w_cadi2}, and this observation motivates us to expand around the relativistic limit.  
Using the results of \aref{app:FD_dist} the equation of state, pseudo-equation of state, and sound speed are written as
\begin{align}
	w = \frac{1}{3} - 2\Delta \cadi^2
	\ , \quad
	\tilde{w} = \frac{1}{3} - 4\Delta \cadi^2
	\ , \quad \text{and} \qquad
	\cadi^2 = \frac{1}{3} - \Delta \cadi^2 
\end{align}
where all of the perturbations are proportional to $m^2/T^2$ and we have used \eref{eq:Dw_Dcadi2}.  
We can similarly expand the perturbation variables around \eref{eq:pert_rel_limit} as\footnote{Explicit calculation using \erefs{app:moments_1}{app:moments_2} reveals that $\delta/3 - \tilde{\Pi} \approx 3 (\delta/3 - \Pi)$ to leading order in $m^2/T^2$.  }
\begin{align}
	& \Pi = \delta/3 - \Delta \Pi
	\ , \quad 
	\tilde{\Pi} = \delta/3 - 3 \Delta \Pi \com \nn
	& \tilde{\theta} = \theta - \Delta \tilde{\theta}
	\ , \quad \text{and} \quad 
	\tilde{\sigma} = \sigma - \Delta \tilde{\sigma}
\end{align}
where the deviations are $O(m^2/T^2)$.  
Making these replacements the collisionless fluid equations \eref{eq:Fluid_Eqns_massive} become
\begin{subequations}\label{eq:Fluid_Eqns_semirel}
\begin{align}
	\dot{\delta} & 
	= - \frac{4}{3} \Bigl( \theta + \frac{1}{2} \dot{h} \Bigr) + 2 \Delta \cadi^2 \Bigl( \theta + \frac{1}{2} \dot{h} \Bigr)  - 3 \frac{\dot{a}}{a} \Bigl( 2 \Delta \cadi^2 \delta - \Delta \Pi \Bigr) \\
	\dot{\theta} & 
	= - 3 \frac{\dot{a}}{a} \Delta c_{\rm adi}^2 \theta + \frac{1}{4} k^2 \delta - k^2 \sigma + \frac{3}{16} k^2 \bigl( 2 \Delta \cadi^2 \delta - 4 \Delta \Pi \bigr) \\
	\dot{\sigma} & 
	= \frac{4}{15} \theta + \bigl( 1 - 3 \Delta \cadi^2 \bigr) \frac{2}{15} \bigl( \dot{h} + 6 \dot{\eta} \bigr) - \frac{3}{5} k \chi - 3 \frac{\dot{a}}{a} \Delta c_{\rm adi}^2 \sigma - \frac{\dot{a}}{a} \Delta \tilde{\sigma} - \frac{4}{15} \Delta \tilde{\theta} 
	\per
\end{align}
\end{subequations}
Here we keep only terms up to linear order in the deviations.  
In summary, a system of free-streaming particle obeys the collisionless fluid equations of \eref{eq:Fluid_Eqns_massive}, and if the particles are semi-relativistic these equations can be approximated as in \eref{eq:Fluid_Eqns_semirel}.  

Now we seek to compare \eref{eq:Fluid_Eqns_semirel} with the generalized fluid equations of \eref{eq:Fluid_Eqns_generalized}.  
To facilitate the comparison we difference the two sets of equations to obtain
\begin{subequations}\label{eq:difference}
\begin{align}
	\dot{\delta} : \ & 
	3 \frac{\dot{a}}{a} \Bigl( \frac{1}{3} - c_{\rm eff}^2 + 2 \Delta \cadi^2 \Bigr) \delta + 12 \Bigl( \frac{\dot{a}}{a} \Bigr)^2 \Bigl( \frac{1}{3} - c_{\rm eff}^2 \Bigr) \frac{\theta}{k^2} - 2 \Delta \cadi^2 \Bigl( \theta + \frac{1}{2} \dot{h} \Bigr) - 3 \frac{\dot{a}}{a} \Delta \Pi \\
	\dot{\theta} : \ & 
	-3 \frac{\dot{a}}{a} \Bigl( \frac{1}{3} - c_{\eff}^2 - \Delta \cadi^2 \Bigr) \theta + \frac{3}{4} \Bigl( c_{\eff}^2 - \frac{1}{3} - \frac{\Delta \cadi^2}{2} \Bigr) k^2 \delta + \frac{3}{4} k^2 \Delta \Pi \\
	\dot{\sigma} : \ & 
	\Bigl( c_{\vis}^2 - \frac{1}{3} \Bigr) \frac{4}{5} \theta + \Bigl( c_{\vis}^2 - \frac{1}{3} + \Delta \cadi^2 \Bigr) \frac{2}{5} \bigl( \dot{h} + 6 \dot{\eta} \bigr) + 3 \frac{\dot{a}}{a} \Delta c_{\rm adi}^2 \sigma + \frac{\dot{a}}{a} \Delta \tilde{\sigma} + \frac{4}{15} \Delta \tilde{\theta} 
	\per
\end{align}
\end{subequations}
Evidently, there is no choice of $\ceff^2$ and $\cvis^2$ that brings the two expressions into the same form, {\it i.e.} causes the three lines of \eref{eq:difference} to vanish.  
The generalized fluid equations thus fail to capture even this minor deviation from the null hypothesis.  

\section{Estimate Deviations from Null Hypothesis}\label{sec:Deviations}

While we have shown that there is no choice of $\ceff^2$ and $\cvis^2$ for which the generalized fluid equations reduce to the collisionless fluid equations, nevertheless, it is reasonable to ask the following question.  
Suppose that the neutrinos have a small mass and are semi-relativistic at the time of recombination.  
This affects the evolution of their density perturbations according to \eref{eq:Fluid_Eqns_semirel} and ultimately impacts the CMB temperature anisotropies.  
However, suppose one (naively) analyzes the observed CMB data using the generalized fluid equations, \eref{eq:Fluid_Eqns_generalized}, which do not capture the physics of the semi-relativistic neutrinos.  
How will the best fit parameters $\ceff^2$ and $\cvis^2$ depend on the neutrino mass?  

Inspecting \eref{eq:difference}, we ask what choice of the phenomenological sound speed and viscosity parameters would give the best agreement between the GFE and collisionless fluid equations.  
Taking $\cvis^2 = 1/3 - \Delta \cadi^2$ causes the gravitational source term to vanish from the equation for $\dot{\sigma}$, and taking $\ceff^2 = 1/3 - \Delta \cadi^2$ causes a number of other terms to exactly or partially cancel.  
This observation suggests that as the neutrinos start becoming semi-relativistic, the sound speed and viscosity will deviate from the null hypothesis, $(\ceff^2, \cvis^2) = (1/3,1/3)$, according to\footnote{A similar identification was employed in the mixed dark matter scenario of \rref{Hu:1998kj}.  } 
\begin{align}\label{eq:Dc_guess}
	\ceff^2 \approx \cadi^2
	\qquad \text{and} \qquad
	\cvis^2 \approx \cadi^2
	\per
\end{align}
While the identification of $\ceff^2$ and $\cvis^2$ with the adiabatic sound speed is not rigorous, we propose here that it quantitatively reflects the correct parametric behavior and order of magnitude of the effect. 
 
It is interesting to note that {\it both} $\ceff^2$ and $\cvis^2$ begin to deviate from $1/3$ as the neutrinos become semi-relativistic.  
This is somewhat surprising, because if we relax only the free streaming assumption, it is possible to describe a relativistic perfect fluid with $(\ceff^2,\cvis^2) = (1/3,0)$ in which only $\cvis^2$ deviates from its value in the null hypothesis.  

One additional comment is in order.  
Whereas $\cadi^2$ is temperature dependent, see \fref{fig:w_cadi2}, the phenomenological parameters $\ceff^2$ and $\cvis^2$ are assumed to be static.  
Thus we should interpret \eref{eq:Dc_guess} to mean that $\ceff^2$ and $\cvis^2$ are derived from a weighted time average of $\cadi^2$ between the epoch of neutrino decoupling and recombination.  
Our analytic approximation does not determine which function will appear in the time averaging.  
However, since $\cadi^2$ decreases monotonically from $1/3$, any arbitrarily weighted time average must satisfy   
\begin{align}
	\Delta \ceff^2 \ , \ \Delta \cvis^2 \ \leq \ \Delta \cadi^2(T_{\rm rec}) \com
\end{align}
where the deviation in the adiabatic sound speed is evaluated at the time of recombination when the neutrino temperature was $T_{\rm rec} \simeq 0.2 \eV$.  
The largest effect of finite neutrino mass on the phenomenological parameters occurs if the inequality is saturated.  
We will make this assumption for determining our upper limits in numerical estimates below.  

Using the analytic expression for $\Delta \cadi^2$ from \eref{eq:Dw_Dcadi2} we estimate 
\begin{align}
	\Delta \ceff^2 \approx \Delta \cvis^2 \lesssim 0.01 \frac{m^2}{T_{\rm rec}^2} 
	\per
\end{align}
The sum of the relic neutrino masses is constrained as in \eref{eq:measure_Neff} using Planck data.  
If the limit is saturated, the neutrinos are in the degenerate regime, and we can take $m \sim 0.08 \eV$ as a reference point.  
For this mass, the anticipated deviation in the sound speed and viscosity parameters at the time of recombination are
\begin{align}\label{eq:active_deviations}
	\Delta \ceff^2 \approx \Delta \cvis^2 \lesssim 0.002 \, \left( \frac{m}{0.08 \eV} \right)^2 \left( \frac{T_{\rm rec}}{0.2 \eV} \right)^{-2} 
	\per
\end{align}
Comparing with \eref{eq:measure_csq}, we see that the expected deviation is smaller than Planck's sensitivity to $\ceff^2$ and $\cvis^2$.  
If the sensitivity to $\ceff^2$ improves by an order of magnitude, the estimate of \eref{eq:active_deviations} suggests that the effect of finite neutrino mass could become relevant.  In that case, a more detailed numerical analysis would be necessary to determine actual constraints..

Next we consider the possibility that the relic neutrino background contains a sub-dominant component of {\rm eV}-scale sterile neutrinos.  
The fact that the neutrinos are sterile, {\it i.e.} not weakly interacting, will not actually be relevant for this discussion.  
Rather, it only matters that they are semi-relativistic and free-streaming at the time of recombination.  
Once again we ask the question:  suppose that the CMB sky generated in this model is studied (naively) using the generalized fluid equations, which do not explicitly account for the sterile neutrino component.  
How will the best fit phenomenological parameters, $\ceff^2$ and $\cvis^2$, depend on the sterile mass and abundance?  

To study this model, one writes down two sets of collisionless fluid equations with each taking the form of \eref{eq:Fluid_Eqns_massive} but labeled by subscripts ``{\rm a}'' for active and ``{\rm s}'' for sterile.  
This significantly complicates the analysis, but we now proceed to argue that one can reduce the system to a single dynamical degree of freedom in the limit where both active and sterile neutrinos are relativistic.  
Since the neutrinos are free-streaming, they only influence the densities of other species ({\it e.g.}, photons) through their gravitational effect on the metric perturbations.  
Einstein's equations, which govern the evolution of the metric perturbations, only depend on the diagonal linear combinations, {\it e.g.} $\bar{\rho}_{\rm a} + \bar{\rho}_{\rm s}$ and $\drho_{\rm a} + \drho_{\rm s}$ (see \rref{Ma:1995ey} for complete expressions).  
Thus, as far as Einstein's equations are concerned, we do not need to know the separate evolution of the active and sterile neutrino perturbation variables, but only their sums are relevant:  
\begin{align}
	& \bar{\rho}_{\nu} = \bar{\rho}_{\rm a} + \bar{\rho}_{\rm s}
	\quad , \quad
	\bar{P}_{\nu} = \bar{P}_{\rm a} + \bar{P}_{\rm s} 
	\quad , \quad
	\drho_{\nu} = \drho_{\rm a} + \drho_{\rm s}
\end{align}
and so on for the other perturbation variables, $\theta_{\nu}$, $\sigma_{\nu}$, etc.
In this way, we can model the combined active and sterile neutrino background as a two-component fluid.  
The corresponding adiabatic sound speed is given by 
\begin{align}\label{eq:cnusq}
	c_{\rm adi,\nu}^2 & = \frac{\dot{\bar{P}}_{\nu}}{\dot{\bar{\rho}}_{\nu}} = \frac{c_{\rm adi,a}^2 (1+w_{\rm a}) \bar{\rho}_{\rm a} + c_{\rm adi,s}^2 (1+w_{\rm s}) \bar{\rho}_{\rm s}}{(1+w_{\rm a}) \bar{\rho}_{\rm a} + (1+w_{\rm s}) \bar{\rho}_{\rm s}} 
\end{align}
where we have used \eref{eq:rhobar_dot}.  
In the subsequent analysis we will assume, as above, that the effective sound speed and viscosity will follow the adiabatic sound speed as the neutrinos become semi-relativistic.  

Note that while the sound speed formula bears a similarity to the baryon-photon fluid, the physics is very different.  
Before recombination the baryons and photons are tightly coupled due to frequent Thompson scattering \cite{Ma:1995ey}.  
Consequently, the baryon perturbation variables tend to track the photon perturbation variables, {\it e.g.} $\theta_{\gamma} \approx \theta_{\rm b}$ and $\sigma_{\gamma} \approx \sigma_{\rm b} \approx 0$, and the single coupled fluid evolves as if it had an adiabatic sound speed given by the analog of \eref{eq:cnusq}.  
In the case of free-streaming neutrinos, on the other hand, the active and sterile perturbations are not directly coupled.  
However, in the relativistic regime, $m_{\rm a} , m_{\rm s} \ll T$, the two sets of Boltzmann equations describing the evolution of the active and sterile neutrinos are reduced to the same form, {\it i.e.} $w_{\rm a} \approx w_{\rm s} \approx 1/3$ and $c_{\rm adi,a}^2 \approx c_{\rm adi,s}^2 \approx 1/3$.  
If the isocurvature modes vanish initially, {\it e.g.} $\theta_{\rm a} \approx \theta_{\rm s}$, then they remain vanishing as long at the both species are ultra-relativistic.  
Consequently the active and sterile neutrino perturbations evolve in the same way, even through they are not directly coupled, and they can be modeled as a single fluid\footnote{One makes a similar reduction when modeling the Standard Model relic neutrino background as a single fluid, even though it is composed of three non-interacting components, corresponding to the three neutrino mass eigenstates.  }.  
Once the sterile neutrinos become non-relativistic, the isocurvature modes will grow, and the two species will start evolving differently.  
Until that time, in the semi-relativistic regime, the sound speed given in \eref{eq:cnusq} is appropriate.  

Further, we assume that the sterile neutrinos have a phase space distribution function of the Fermi-Dirac form with the same temperature as the active neutrinos but a different overall normalization:  
\begin{align}\label{eq:f0_sterile}
	f_{0, {\rm a}}(q) = \frac{g}{(2\pi)^3} \frac{1}{e^{q/aT} + 1 }
	\qquad \text{and} \qquad
	f_{0, {\rm s}}(q) = \alpha \, f_{0,{\rm a}}(q) 
	\per
\end{align}
The proportionality constant $\alpha$ controls the relative number densities, $\bar{n}_{\rm s} = \alpha \, \bar{n}_{\rm a}$.  
In the relativistic limit this proportionality implies $\bar{\rho}_{\rm s} / \bar{\rho}_{\rm a} \approx \alpha$, and \eref{eq:Neff_and_summnu} gives
\begin{align}
	\Delta N_{\rm eff} \approx \frac{8}{7} \left( \frac{11}{4} \right)^{4/3} \frac{\bar{\rho}_{\rm s}}{\rho_{\gamma}} \approx 3 \alpha 
	\per
\end{align}
In the non-relativistic limit the proportionality implies
\begin{align}
	\Delta \sum m_{\nu} \approx m_{\rm s} \frac{\bar{n}_{\rm s}}{\bar{n}_{\rm a}} \approx \alpha \, m_{\rm s} \per
\end{align}
The effective number of neutrinos is measured with an error of $\dN_{\rm eff} \approx 0.4$, see \eref{eq:measure_Neff}, which implies $\bar{\rho}_{\rm s} / \bar{\rho}_{\rm a} = \alpha \lesssim \dN_{\rm eff} / 3 \approx 0.1$.  
Similarly, imposing the bound on $\sum m_{\nu}$ implies $m_{\rm s} \lesssim (0.2 \eV)  / \alpha \approx 2 \eV$ for $\alpha \approx 0.1$.  

\begin{figure}[t]
\hspace{0pt}
\vspace{-0.0cm}
\begin{center}
\includegraphics[width=0.65\textwidth]{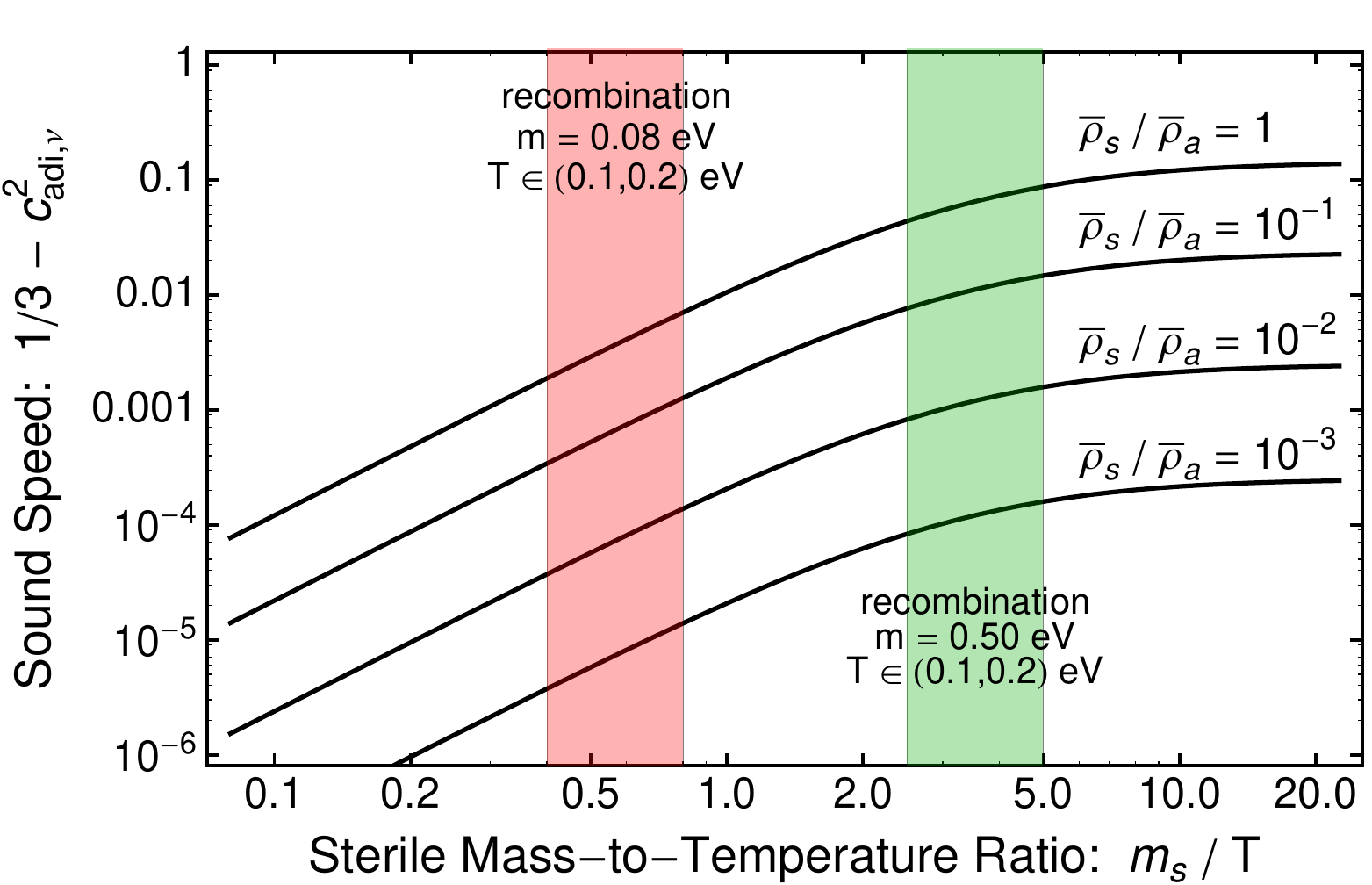} 
\end{center}
\caption{\label{fig:sound_speed}  
The relic neutrino adiabatic sound speed from \eref{eq:cnusq}.   Varying the sterile neutrino mass, $m_{\rm s}$, affects the equation of state and sound speed, $w_{\rm s}$ and $c_{\rm adi,s}^2$, which appear in \eref{eq:cnusq}.  The three lines correspond to different values of the sterile-to-active energy ratios, $\bar{\rho}_{\rm s} / \bar{\rho}_{\rm a} = 1, 10^{-1}, 10^{-2}$, and $10^{-3}$ from top to bottom, which is a proxy for $\Delta N_{\rm eff}/3$.  
}
\end{figure}

\begin{figure}[t]
\hspace{0pt}
\vspace{-0.0cm}
\begin{center}
\includegraphics[width=0.65\textwidth]{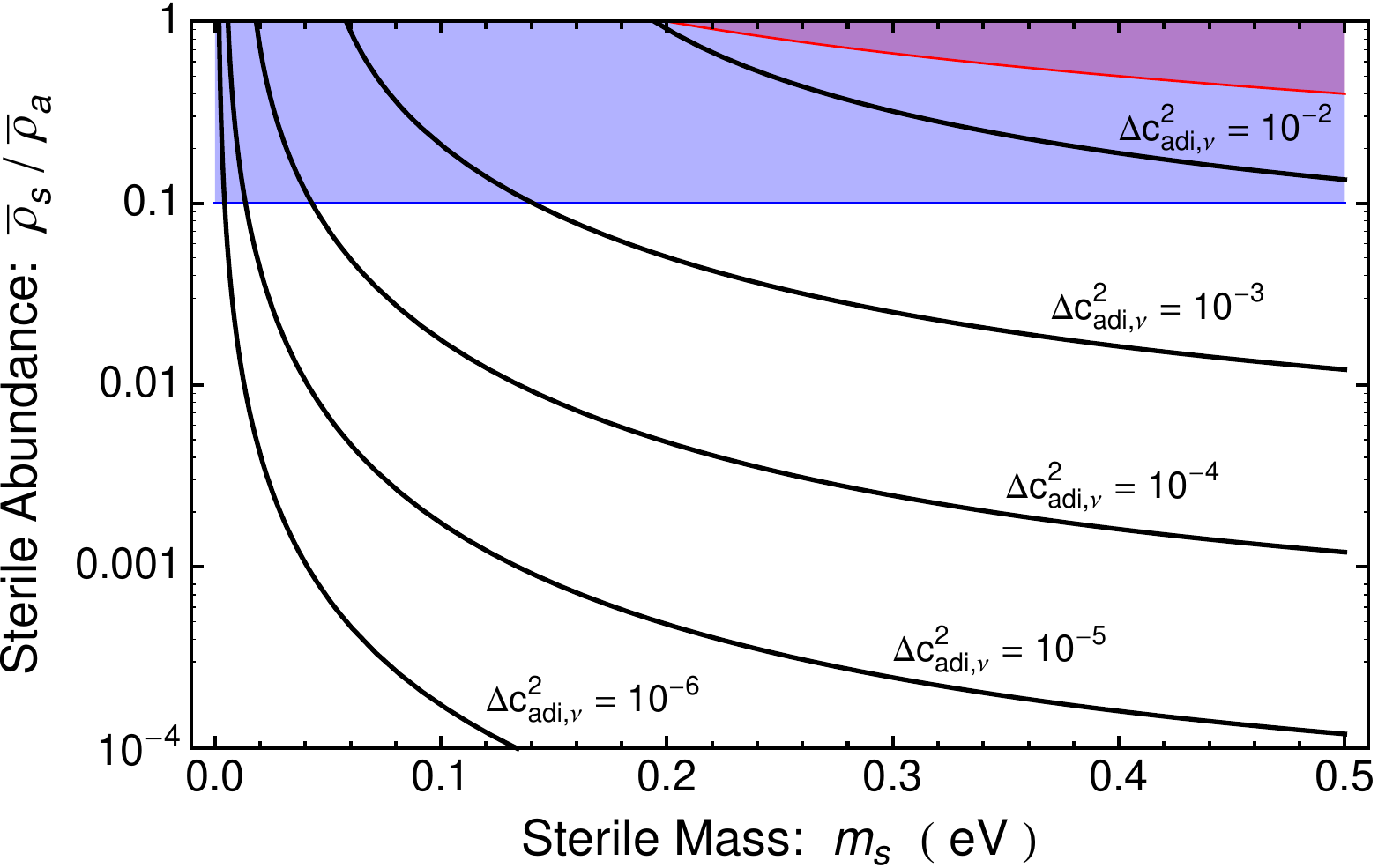} 
\end{center}
\caption{\label{fig:pspace}  
The relic neutrino adiabatic sound speed from \eref{eq:cnusq}.  The figure is calculated in the same way as \fref{fig:sound_speed}, but we hold $T = 0.2 \eV$ fixed.  Curves of constant $\Delta c_{\rm adi,\nu}^2 = 1/3 - c_{\rm adi,\nu}^2$ are shown.  The red shaded region is excluded by the bound on $\sum m_{\nu}$, and the blue shaded region is excluded by the bound on $\Delta N_{\rm eff}$.  
}
\end{figure}

We evaluate the relic neutrino adiabatic sound speed using \eref{eq:cnusq}.  
The active neutrinos are still relativistic at recombination, and we can set $w_{\rm a} \approx c_{\rm adi,a}^2 \approx 1/3$.  
The sterile equation of state and sound speed, $w_{\rm s}$ and $c_{\rm adi,s}^2$, are calculated from \eref{eq:f0_sterile}; they vary with the sterile neutrino mass $m_{\rm s}$ as shown in \fref{fig:w_cadi2}.  
Figures~\ref{fig:sound_speed}~and~\ref{fig:pspace} show how the sound speed deviation $\Delta c_{\rm adi,\nu}^2 = 1/3 - c_{\rm adi,\nu}^2$ varies with the sterile neutrino mass $m_{\rm s}$ and relative abundance $\bar{\rho}_{\rm s} / \bar{\rho}_{\rm a}$.  
If the sterile neutrino is sufficiently light, then it is still relativistic at recombination, and its effect on the sound speed is small.  
Similarly, if the relative sterile abundance is small, $\alpha = \bar{\rho}_{\rm s} / \bar{\rho}_{\rm a} \ll 1$, then it also has a suppressed impact on the sound speed.  
As the sterile neutrino mass is increased, the sound speed begins to deviate further from the null hypothesis value, $c_{\rm adi,\nu}^2 =1/3$.  
However, if the mass is so large that the sterile neutrino is non-relativistic at recombination, then the approximations used in our calculation are no longer valid, which is why we cut off Figures~\ref{fig:sound_speed} at $m_s/T \approx 20$.  
\fref{fig:pspace} also indicates the parameter space that is excluded by bounds on $\sum m_{\nu}$ and $\Delta N_{\rm eff}$ as the red and blue shaded regions, respectively.  
Focusing on $m_{\rm s} \approx T_{\nu} \simeq 0.2 \eV$ we see that the sound speed can deviate from the null hypothesis by as much as $O(10^{-3})$ before running into the bound on $N_{\rm eff}$.  

\section{Conclusion and Discussion}\label{sec:Discussion}

Generalized fluid equations (GFEs) provide a phenomenological formalism for testing the relic neutrino ``null hypothesis,'' {\it i.e.} that the neutrinos are both relativistic and free-streaming in the epoch prior to recombination.  
This formalism has two key advantages:  it requires only minimal modifications to the fluid equations, which are easily implemented in a numerical Boltzmann solver; and these modifications are captured by just two parameters, the effective sound speed $\ceff^2$ and the viscosity $\cvis^2$, which quantify deviations from the null hypothesis.  However, as we have demonstrated in Sec. 2, where we consider the effects of finite neutrino masses, specific microphysical models that deviate from the null hypothesis cannot always be accommodated into the GFE formalism.  

One can nevertheless investigate how the presence of a finite neutrino mass would affect the best fit values of $\ceff^2$ and $\cvis^2$ if the CMB sky were analyzed using a GFE analysis.  
In \sref{sec:Deviations} we propose that one can use the adiabatic sound speed at recombination, $\cadi^2(T_{\rm rec})$, to gauge the magnitude of deviations in $\ceff^2$ and $\cvis^2$ from $1/3$, the null hypothesis prediction.  
Taking this as our measure, we estimate that a value of the neutrino mass saturating the Planck limit, $m \simeq 0.08 \eV$, could induce a deviation in the effective sound speed and viscosity by as much as $0.2 \%$.  
If the relic neutrino background contains a sterile component, similar estimates suggest that deviations could be as large as $\Delta \ceff^2 \approx \Delta \cvis^2 \approx 10^{-3}$ if Plank's limits on $\Delta N_{\rm eff}$ and $\sum m_{\nu}$ are saturated.  
Since Planck's error bars on the phenomenological parameters are relatively large, $\delta \ceff^2 \simeq 0.0059$ and $\delta \cvis^2 \simeq 0.037$ respectively, the effect of finite neutrino mass is currently imperceptible.  
However, if the next generation of CMB telescopes achieve an order of magnitude improvement in sensitivity to the GFE parameters, then our estimates suggest that the effect of finite neutrino mass cannot be neglected, and analytical and phenomenological approximations will need to be supplemented by detailed numerical estimates.   

In this regard we conclude by noting that our numerical estimates of \sref{sec:Deviations} rely on the plausible argument that the best fit values of $\ceff^2$ and $\cvis^2$ will begin to deviate from $1/3$ in the same way as $\cadi^2$.  
If experimental sensitivities improve sufficiently one could test this ansatz in detail using a numerical Boltzmann solver and the following algorithm\footnote{The algorithm we outline here is similar in spirit to the approach taken by Refs.~\cite{Jacques:2013xr, Grohs:2014rea} to infer the effect of neutrino mass on $N_{\rm eff}$.  }:  for a particular neutrino mass solve the full Boltzmann hierarchy, \eref{app:BH}, to generate realizations of the CMB sky; then for a particular $(\ceff^2, \cvis^2)$ solve the generalized fluid equations \eref{eq:Fluid_Eqns_generalized};  using MCMC techniques find the values of $(\ceff^2, \cvis^2)$ that best fit the sky generated from the Boltzmann hierarchy.   

\begin{acknowledgments}
We are grateful for discussions with Sean Bryan, Daniel Grin, Wayne Hu, and Marius Millea, and we are grateful to Cecilia Lunardini for collaboration in the initial stages of this project.  
This work was supported at Arizona State University by the Department of Energy under Grant No.\ DE-SC0008016 and the National Science Foundation under Grant No.\ PHY-1205745.  
This work was supported at the University of Chicago by the Kavli Institute for Cosmological Physics through grant NSF PHY-1125897 and an endowment from the Kavli Foundation and its founder Fred Kavli.  
\end{acknowledgments}

\appendix

\section{Derivation of Fluid Equations}\label{app:FluidEqns}

In this appendix we derive the fluid equations from the Boltzmann hierarchy.  
Our notation mostly follows Ma \& Bertschinger \cite{Ma:1995ey}:  $a(\tau)$ is the scale factor, $\ud \tau = \ud t / a(\tau)$ is the conformal time, $\ud \xvec = \ud \rvec / a(\tau)$ is the comoving coordinate, $\kvec$ is the corresponding wave vector, $\qvec = a(\tau) \pvec$ is the comoving momentum, and $\ebar(\tau) = \sqrt{q^2 + a(\tau)^2 m^2}$ is $a(\tau)$ times the proper energy measured by a comoving observer.  

The phase space distribution function is written as 
\begin{align}
	f(\kvec,\qvec,\tau) & = f_{0}(q,\tau) \bigl( 1 + \Psi(\kvec,\qvec,\tau) \bigr) \per
\end{align}
For freely streaming particles, $f$ satisfies the collisionless Boltzmann equation.  
For the homogenous term\footnote{The analysis in this appendix does not assume any specific form for $f_{0}(q,\tau)$.  } this is simply $\partial f_{0}(q,\tau) / \partial \tau = 0$, and the perturbations satisfy 
\begin{align}
	\frac{\partial \Psi}{\partial \tau} + i \frac{qk}{\ebar} (\khat \cdot \qhat) \Psi + \frac{d \ln f_{0}}{d \ln q} \left( \dot{\eta} - \frac{\dot{h} + 6 \dot{\eta}}{2} (\khat \cdot \qhat)^2 \right) = 0 \com
\end{align}
which has been written here in synchronous gauge (with $\eta(\kvec,\tau)$ and $h(\kvec,\tau)$ the metric perturbations).  
The perturbation is decomposed onto the Legendre polynomials as 
\begin{align}
	\Psi(\kvec,\qvec,\tau) & = \sum_{l=0}^{\infty} (-i)^{l} (2l+1) \, \Psi_{l}(\kvec,q,\tau) \, P_{l}(\mu) 
\end{align}
with $\mu = \khat \cdot \qhat$.  
The Boltzmann equation resolves to the set of coupled, first-order differential equations
\begin{subequations}\label{app:BH}
\begin{align}
	\dot{\Psi}_{0} & = - \frac{qk}{\ebar} \Psi_{1} + \frac{1}{6} \frac{d \ln f_{0}}{d \ln q} \dot{h} \label{app:BH0} \\
	\dot{\Psi}_{1} & = \frac{qk}{3\ebar} \Bigl( \Psi_{0} - 2 \Psi_{2} \Bigr) \label{app:BH1} \\
	\dot{\Psi}_{2} & = \frac{qk}{5\ebar} \Bigl( 2 \Psi_{1} - 3 \Psi_{3} \Bigr) - \frac{1}{15} \frac{d \ln f_{0}}{d \ln q} \bigl( \dot{h} + 6 \dot{\eta} \bigr) \label{app:BH2} \\
	\dot{\Psi}_{l} & = \frac{1}{2l+1} \frac{qk}{\ebar} \Bigl( l \Psi_{l-1} - (l+1) \Psi_{l+1} \Bigr) \qquad \text{for  $l \geq 3$} \label{app:BHl}
	\com
\end{align}
\end{subequations}
which are collectively known as the Boltzmann hierarchy.  

If one is not interested in the momentum dependence of the perturbations, it would seem that the problem is simplified by integrating \eref{app:BH} over $q$.  
In the case of massless particles ($\ebar = q$) one can identify a new dynamical variable $F_{l}(\kvec,\tau) \propto \int_{0}^{\infty} \! q^2 \ud q \, q f_{0}(q,\tau) \Psi_{l}(\kvec,q,\tau)$ for each original $\Psi_{l}$, and in fact, the problem is simplified.  
However in the massive case ($\ebar \neq q$) the number of dynamical variables increases.  
For instance, \eref{app:BH0} gives the evolution of $A_{0} = \int \ebar \Psi_{0}$ in terms of $B_{1} = \int q \Psi_{1}$ (written schematically), but \eref{app:BH1} gives the evolution of $B_{1}$ in terms of $C_{0} = \int (q^2/\ebar) \Psi_{0}$.  
This second moment of $\Psi_{0}$ requires its own evolution equation, and thus it is typically easier to solve \eref{app:BH} directly.  
Nevertheless, the first few equations obtained by integrating \eref{app:BH} correspond to the familiar fluid equations, and we now proceed to derive them.  

First we define the spatially averaged energy density, pressure, and pseudo-pressure:  
\begin{align}
	\bar{\rho}(\tau) & = 4\pi a(\tau)^{-4} \int_{0}^{\infty} \! q^2 \ud q \, f_{0}(q,\tau) \, \ebar(q,\tau) \label{eq:f0_to_rhobar} \\
	\bar{P}(\tau) & = 4\pi a(\tau)^{-4} \int_{0}^{\infty} \! q^2 \ud q \, f_{0}(q,\tau) \, \frac{q^2}{3\ebar(q,\tau)} \label{eq:f0_to_Pbar} \\
	\tilde{P}(\tau) & = 4 \pi a(\tau)^{-4} \int_{0}^{\infty} \! q^2 \ud q \, f_{0}(q,\tau) \, \frac{q^4}{3\ebar(q,\tau)^3} \label{eq:f0_to_tildePbar} 
	\per
\end{align}
For freely streaming particles ($\partial f_0 / \partial \tau = 0$), the energy density satisfies the homogenous continuity equation
\begin{align}\label{eq:rhobar_dot}
	\dot{\bar{\rho}}(\tau) = - 3\frac{\dot{a}}{a} \bigl( \bar{\rho} + \bar{P} \bigr) \com
\end{align}
and the pressure satisfies
\begin{align}
	\dot{\bar{P}}(\tau) = - \frac{\dot{a}}{a} \bigl( 5 \bar{P} - \tilde{P} \bigr) \per
\end{align}
We define the equation of state, pseudo-equation of state, and adiabatic sound speed,
\begin{align}\label{eq:w_tildew_cadi2_def}
	w(\tau) = \frac{\bar{P}(\tau)}{\bar{\rho}(\tau)} 
	\ , \quad 
	\tilde{w}(\tau) = \frac{\tilde{P}(\tau)}{\bar{\rho}(\tau)} 
	\ , \quad \text{and} \quad
	c_{\rm adi}^2(\tau) = \frac{\dot{\bar{P}}(\tau)}{\dot{\bar{\rho}}(\tau)}
	\per
\end{align}
They obey the useful relations
\begin{align}
	\frac{\dot{w}}{1+w} = 3 \frac{\dot{a}}{a} \bigl( w - c_{\rm adi}^2 \bigr)
	\quad \text{and} \quad
	c_{\rm adi}^2 = \frac{5 w - \tilde{w}}{3(1 + w)} \per
\end{align}
Next we define a few of the lower order moments of $\Psi_{l}$ as\footnote{Here our notation diverges from that of Ma \& Bertschinger \cite{Ma:1995ey}.}
\begin{subequations}\label{app:moments_1}
\begin{align}
	\drho(\kvec,\tau) & 
	= 4\pi a(\tau)^{-4} \int_{0}^{\infty} \! q^2 \ud q \, f_{0}(q,\tau) \Psi_{0}(\kvec,q,\tau) \, \ebar(q,\tau) \\
	\dP(\kvec,\tau) & 
	= 4\pi a(\tau)^{-4} \int_{0}^{\infty} \! q^2 \ud q \, f_{0}(q,\tau) \Psi_{0}(\kvec,q,\tau) \, \frac{q^2}{3\ebar(q,\tau)} \\
	\dPtilde(\kvec,\tau) & 
	= 4\pi a(\tau)^{-4} \int_{0}^{\infty} \! q^2 \ud q \, f_{0}(q,\tau) \Psi_{0}(\kvec,q,\tau) \, \frac{q^4}{3\ebar(q,\tau)^3} \\
	\dQ(\kvec,\tau) & 
	= 4\pi a(\tau)^{-4} \int_{0}^{\infty} \! q^2 \ud q \, f_{0}(q,\tau) \Psi_{1}(\kvec,q,\tau) \, q k \\
	\dQtilde(\kvec,\tau) & 
	= 4\pi a(\tau)^{-4} \int_{0}^{\infty} \! q^2 \ud q \, f_{0}(q,\tau) \Psi_{1}(\kvec,q,\tau) \, \frac{q^3 k}{\ebar(q,\tau)^2} \\
	\dS(\kvec,\tau) & 
	= 4\pi a(\tau)^{-4} \int_{0}^{\infty} \! q^2 \ud q \, f_{0}(q,\tau) \Psi_{2}(\kvec,q,\tau) \, \frac{2q^2}{3 \ebar(q,\tau)} \\
	\dStilde(\kvec,\tau) & 
	= 4\pi a(\tau)^{-4} \int_{0}^{\infty} \! q^2 \ud q \, f_{0}(q,\tau) \Psi_{2}(\kvec,q,\tau) \, \frac{2q^4}{3\ebar(q,\tau)^3} \\
	\dC(\kvec,\tau) & 
	= 4\pi a(\tau)^{-4} \int_{0}^{\infty} \! q^2 \ud q \, f_{0}(q,\tau) \Psi_{3}(\kvec,q,\tau) \, \frac{2q^3}{3\ebar(q,\tau)^2} 
	\per
\end{align}
\end{subequations}
These correspond to perturbations in the energy density $\drho$, pressure $\dP$, pseudo-pressure $\dPtilde$, energy flux $\dQ$, anisotropic stress $\dS$, etc.  
We can also write
\begin{subequations}\label{app:moments_2}
\begin{align}
	\drho(\kvec,\tau) & = \bar{\rho}(\tau) \, \delta(\kvec,\tau) \\ 
	\dP(\kvec,\tau) & = \bar{\rho}(\tau) \, \Pi(\kvec,\tau) \\
	\dPtilde(\kvec,\tau) & = \bar{\rho}(\tau) \, \tilde{\Pi}(\kvec,\tau) \\
	\dQ(\kvec,\tau) & = \bigl( 1+w(\tau) \bigr) \, \bar{\rho}(\tau) \, \theta(\kvec,\tau) \\
	\dQtilde(\kvec,\tau) & = \bigl( 1+w(\tau) \bigr) \, \bar{\rho}(\tau) \, \tilde{\theta}(\kvec,\tau) \\
	\dS(\kvec,\tau) & = \bigl( 1+w(\tau) \bigr) \, \bar{\rho}(\tau) \, \sigma(\kvec,\tau) \\
	\dStilde(\kvec,\tau) & = \bigl( 1+w(\tau) \bigr) \, \bar{\rho}(\tau) \, \tilde{\sigma}(\kvec,\tau) \\
	\dC(\kvec,\tau) & = \bigl( 1+w(\tau) \bigr) \, \bar{\rho}(\tau) \, \chi(\kvec,\tau) \com
\end{align}
\end{subequations}
which defines the dimensionless perturbation variables $\delta$, $\Pi$, etc.  
For massless particles ($\ebar = q$) we have $w = \tilde{w} = c_{\rm adi}^2 = 1/3$, and the higher order moments simplify to the lower order ones, {\it e.g.} $\dQtilde = \dQ$, $\tilde{\theta} = \theta$, $\tilde{P} = P$, and so on.  

Finally we are prepared to derive the fluid equations from the Boltzmann hierarchy.  
Taking the appropriately-weighted momentum integral of \eref{app:BH0} leads to the inhomogenous continuity equation, which can be written in three equivalent forms:
\begin{subequations}\label{app:Ceqn}
\begin{align}
	\dot{\drho} & 
	= - \dQ - \frac{1}{2} \dot{h} \, \bigl( \bar{\rho} + \bar{P} \bigr) - 3 \frac{\dot{a}}{a} \bigl(\drho + \dP \bigr) \\
	\dot{\delta} & 
	= - (1+w) \Bigl( \theta + \frac{1}{2} \dot{h} \Bigr) + 3 \frac{\dot{a}}{a} \Bigl( w \delta - \Pi \Bigr) \\
	\left( \frac{\delta}{1+w} \right)^{\! \! \cdot} & 
	= - \Bigl( \theta + \frac{1}{2} \dot{h} \Bigr) + 3 \frac{\dot{a}}{a} \left( c_{\rm adi}^2 \frac{\delta}{1+w} - \frac{\Pi}{1+w} \right)
	\per
\end{align}
\end{subequations}
The relation $\Pi = (\dP/\drho) \delta$ puts the second equation into a more familiar form.  
Using a different weighting in the momentum integral yields, 
\begin{subequations}\label{app:Peqn}
\begin{align}
	\dot{\dP} & 
	= -5 \frac{\dot{a}}{a} \dP + \frac{\dot{a}}{a} \dPtilde - \frac{1}{3} \dQtilde - \frac{1}{6} \dot{h} \bigl( 5 \bar{P} - \tilde{P} \bigr) \\
	\dot{\Pi} & 
	= - 3 \frac{\dot{a}}{a} \Bigl( \frac{1}{3} - w \Bigr) \Pi + \frac{\dot{a}}{a} \bigl( \tilde{\Pi} - \Pi \bigr) - \frac{1}{3} (1+w) \tilde{\theta} - \frac{1}{6} \dot{h} \bigl( 5 w - \tilde{w} \bigr) \\
	\left( \frac{\Pi}{1+w} \right)^{\! \! \cdot} & 
	= - 3 \frac{\dot{a}}{a} \Bigl( \frac{1}{3} - c_{\rm adi}^2 \Bigr) \frac{\Pi}{1+w} + \frac{\dot{a}}{a} \frac{\tilde{\Pi} - \Pi}{1+w} - \frac{1}{3} \tilde{\theta} - \frac{1}{2} c_{\rm adi}^2 \dot{h}
	\com
\end{align}
\end{subequations}
which gives the evolution of the momentum perturbation.  
Integrating \eref{app:BH1} leads to the Euler equation, 
\begin{subequations}\label{app:Eeqn}
\begin{align}
	\dot{\dQ} & 
	= - 4 \frac{\dot{a}}{a} \dQ + k^2 \dP - k^2 \dS \\
	\dot{\theta} & 
	= - 3 \frac{\dot{a}}{a} \Bigl( \frac{1}{3}- c_{\rm adi}^2 \Bigr) \theta + k^2 \frac{\Pi}{1 + w} - k^2 \sigma \com
\end{align}
\end{subequations}
and integrating \eref{app:BH2} gives the shear equation, 
\begin{subequations}\label{app:Seqn}
\begin{align}
	\dot{\dS} & 
	= - 5 \frac{\dot{a}}{a} \dS + \frac{\dot{a}}{a} \dStilde + \frac{4}{15} \dQtilde + \frac{2}{15} ( \dot{h} + 6 \dot{\eta} ) \bigl( 5 \bar{P} - \tilde{P} \bigr) + \frac{3}{5} k \dC \\
	\dot{\sigma} & 
	= - 3 \frac{\dot{a}}{a} \Bigl( \frac{1}{3} - c_{\rm adi}^2 \Bigr) \sigma + \frac{\dot{a}}{a} \bigl( \tilde{\sigma} - \sigma \bigr) + \frac{4}{15} \tilde{\theta} + \frac{2}{5} c_{\rm adi}^2 ( \dot{h} + 6 \dot{\eta} ) - \frac{3}{5} k \chi 
	\per
\end{align}
\end{subequations}
Equations (\ref{app:Ceqn})--(\ref{app:Seqn}) do not form a closed system, since the evolution of $\tilde{\Pi}$, $\tilde{\theta}$, $\tilde{\sigma}$, and $\chi$ are undetermined.  

\section{Fermi-Dirac Distribution}\label{app:FD_dist}

For the relic neutrinos, which decoupled while they were relativistic, $f_{0}(q,\tau)$ maintains the Fermi-Dirac distribution
\begin{align}
	f_{0} = \frac{g}{(2\pi)^3} \frac{1}{e^{q/aT} + 1 }
\end{align}
where $aT = a_0 T_0$ is independent of $\tau$ and $g = 6$ counts the two spin and three flavor degrees of freedom.  
The energy density, pressure, and pseudo-pressure are calculated from Eqs.~(\ref{eq:f0_to_rhobar}),~(\ref{eq:f0_to_Pbar}),~and~(\ref{eq:f0_to_tildePbar}) with $\epsilon = \sqrt{q^2 + a(\tau)^2 m^2}$.  
In the limit $m^2 / T^2 \ll 1$ the integrals can be evaluated analytically, and we find 
\begin{align}
	\bar{\rho}(\tau) & \approx \frac{7}{240} g \pi^2 T^4 + \frac{g}{48} m^2 T^2 \\
	\bar{P}(\tau) & \approx \frac{7}{720} g \pi^2 T^4 - \frac{g}{144} m^2 T^2 \\
	\tilde{P}(\tau) & \approx \frac{7}{720} g \pi^2 T^4 - \frac{g}{48} m^2 T^2 
\end{align}
up to terms that are $O(m^4)$.  
The equation of state, pseudo-equation of state, and adiabatic sound speed are calculated using \eref{eq:w_tildew_cadi2_def}.  
The exact expressions, determined numerically, are shown in \fref{fig:w_cadi2}.  
In the limit $m^2 / T^2 \ll 1$ we can approximate
\begin{align}
	w & \approx \frac{1}{3} - \frac{10}{21 \pi^2} \frac{m^2}{T^2} \\
	\tilde{w} & \approx \frac{1}{3} - \frac{20}{21 \pi^2} \frac{m^2}{T^2} \\
	\cadi^2 & \approx \frac{1}{3} - \frac{5}{21 \pi^2} \frac{m^2}{T^2} \com
\end{align}
up to terms of order $O(m^4 / T^4)$.  
These expressions give \eref{eq:Dw_Dcadi2}.


\providecommand{\href}[2]{#2}\begingroup\raggedright\endgroup

\end{document}